Relative rates of fluid advection, elemental diffusion and replacement govern reaction front patterns

Koehn, Daniel (1); Piazolo, Sandra (2); Beaudoin, Nicolas E. (3); Kelka, Ulrich (4); Spruženiece, Liene (5); Christine V. Putnis (6, 7); Toussaint, Renaud (8, 9)

(1) GeoZentrum Nordbayern, University Erlangen-Nuremberg, Schlossgarten 5, 91054, Erlangen, Germany (daniel.koehn@fau.de)
(2) School of Earth and Environment, Institute of Geophysics and Tectonics, The University of Leeds, Leeds, LS2 9JT, UK
(3) Universite de Pau et des Pays de l'Adour, E2S UPPA, LFCR, Pau, France
(4) CSIRO - Deep Earth Imaging, 26 Dick Perry Ave, Kensington WA 6151, Australia
(5) RWTH Aachen University, Structural Geology, Tectonics and Geomechanics, Lochnerstraße 4-20, Aachen, Germany
(6) Institut für Mineralogie, University of Münster, Corrensstrasse 24, 48149 Münster, Germany
(7) School of Molecular and Life Science, Curtin University, Perth 6845, Australia
(8) Institut de Physique du Globe de Strasbourg, UMR 7516, Université de Strasbourg/EOST, CNRS, 5 rue René Descartes, 67084 Strasbourg Cedex, France.
(9) SFF PoreLab, the Njord Centre, Department of Physics, University of Oslo, P.O. Box 1048 Blindern, NO-0316 Oslo, Norway

Highlights

- Reactive fluid infiltration into granular rocks produces very variable reaction front roughness

- Reaction front roughness is supressed by fast reactions

- Reaction patterns mimic microstructure best with advective transport and slow reaction

- We present a diagram of resulting patterns according to Péclet and Damköhler numbers

Abstract

Replacement reactions during fluid infiltration into porous media, rocks and buildings are known to have important implications for reservoir development, ore formation as well as weathering. Natural observations and experiments have shown that in such systems the shape of reaction fronts can vary significantly ranging from smooth, rough to highly irregular. It remains unclear what process-related knowledge can be derived from these reaction front patterns. In this contribution we show a numerical approach to test the effect of relative rates of advection, diffusion and reaction on the development of reaction fronts patterns in granular aggregates with permeable grain boundaries. The numerical model takes (i) fluid infiltration along permeable grain boundaries, (ii) reactions and (iii) elemental diffusion into account. We monitor the change in element concentration within the fluid, while reactions occur at a pre-defined rate as a function of the local fluid concentration. In non-dimensional phase space using Péclet and Damköhler numbers, results show that there are no rough fronts without advection (Péclet<70) nor if the reaction is too fast (Damköhler>$10^{-3}$). As advection becomes more dominant and reaction slower, roughness develops across several grains with a full microstructure mimicking replacement in the most extreme cases. The reaction front patterns show an increase in roughness with increasing Péclet number from Péclet 10 to 100 but then a decrease in roughness towards higher Péclet numbers controlled by the Damköhler number. Our results indicate that reaction rates are crucial for pattern formation and that the shape of reaction fronts is only partly due to the underlying transport mechanism.

1. Introduction

Fluid infiltration, material transport and related reactions induce mineralogical changes that can dramatically modify the physiochemical properties of rocks affecting their mechanical and hydrodynamic properties (Jamtveit et al., 2000; Putnis and Austrheim, 2010). Incomplete element and mineralogical redistribution are both preserved in the rock record in the form of chemical reaction fronts – the more or less localized interface between reacted and unreacted material. Such fronts control geochemical exchange between the atmosphere, hydrosphere and the geosphere with importance for weathering at the Earth's surface in rocks as well as building stones (Kondratiuk et al., 2017; Ruiz-Agudo et al., 2016) and diagenesis. Understanding reaction fronts also has strong fundamental implications to reconstruct large-scale geodynamic histories based on the occurrence of prograde and retrograde metamorphic reactions that include fluids (Austrheim,1987; Ague, 2003; Centrella et al, 2016; Plümper et al., 2017) as well as retrogressive reactions when buried rocks are exhumed (Rudge et al., 2010; Yardley and Cleverley, 2013). Furthermore, changes due to fluid-rock interaction are of importance for the prediction of reservoir characteristics (e.g. Rochelle et al., 2004), the understanding of geothermal systems including their scaling and the development of mineral deposits (Merino and Canals, 2011). Reaction fronts that are linked to fluid-mediated replacement reactions (fig. 1) have been shown to be common in the rock record (Putnis, 2009). Such reactions require the presence of a fluid in chemical disequilibrium with the surrounding minerals. Fluids in chemical disequilibrium need to be transported to the site of reaction, hence such fluids need pathways to infiltrate the system at a certain rate (Jamtveit et al., 2009; Putnis and Austrheim, 2010; Ulven et al., 2014). Transport occurs within the fluid as well as in the solid where chemical constituents are moving according to both advective and diffusive laws. At the same time, the chemical constituents needed for the replacement reaction must be present allowing the existing phase to dissolve, the interfacial mineral-fluid boundary layer to become supersaturated and a new more stable phase to grow (Ruiz-Agudo et al., 2014). Fluid transport, reaction and diffusion each have a certain rate, that may be all interrelated. Conceptually, different rates should result in different rates of reaction front progression but also in different reaction front patterns with differences in chemical, isotopic and trace-element signatures (Centrella et al., 2016).

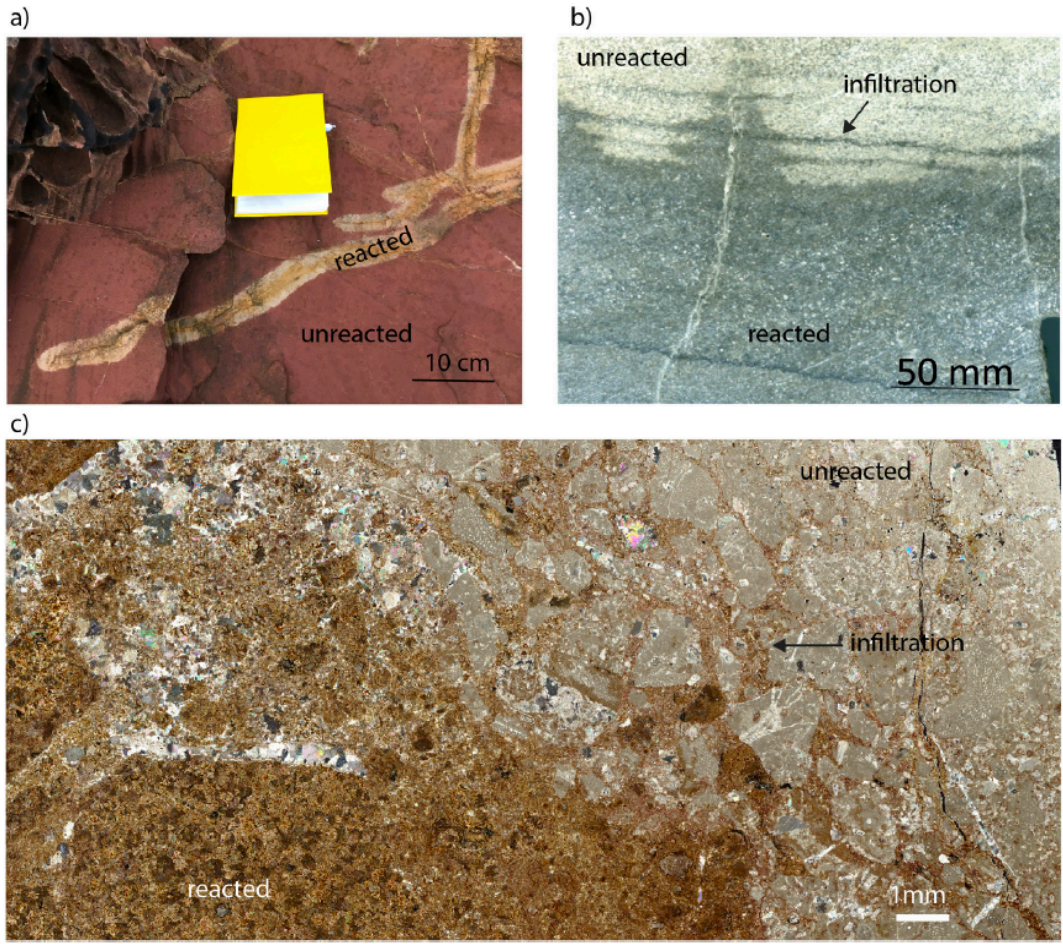

Figure 1. Two examples of reaction textures in natural examples that illustrate the importance of advection. a) Bleaching reactions around a fracture network in red sandstone on the Isle of Cumbrae in Scotland. The reacting fluid clearly came along a fracture network and then affected an area around the fractures. b) Reaction front from the Parmozany mine in Poland. The light top area in the picture is composed of porous dolomite whereas the dark material is composed of very dense dolomite with ore minerals at the bottom of the section. In this case the front is very rough and follows a fracture network upwards and then infiltrates in bedding planes and stylolites sideways. Even though the pattern shows a reaction it does look like the image of an infiltrating fluid. c) Dolomitization front details with dolomitized material on the left hand side in a brown color. The sample is from a zone of fluid driven dolomitization with associated ore deposits in the Oropesa Ranges near Benicassim, Spain. On the sub-

centimetre scale the mineralization front is rough with infiltration along grain-boundaries or zones of smaller grains.

Thus, if we can link the pattern of the reaction front to the relative rates of the three main processes involved, we can use the rock record directly to determine these.

If the reaction only takes place around a fracture in the rock one can of course assume that fluid flow along a fracture network was responsible for the reaction (fig. 1a). However, reaction fronts of large bodies are quite often smooth on the larger (meter to deca-meter) scale and seem to have preserved a pattern that indicates, at least locally, important fluid infiltration along fractures, bedding planes or grain boundaries (fig. 1b). It is not clear how this "fluid-flow" or "infiltration" pattern (fig. 1b) is so clearly preserved in a reaction.

Reactants, i.e. the chemical constituents that can trigger reactions in rocks, can enter the system by two main transport mechanisms: advection and diffusion (Jamtveit and Meakin, 1999). Diffusion takes place where the concentration of the chemical constituent changes along the chemical gradient. This process is relatively slow and scales non-linearly with the square root of time (Jamtveit and Meakin, 1999). Therefore, it is either important on the very small scale or over very long (geological) timescales. Advection on the other hand involves fluid-flow through the system either by wetting a dry rock or through convection cells driven by thermal or salinity induced density contrasts, for example in geothermal systems (Lipsey et al., 2016). The reactants are then brought in with the fluid and the timescale of this process depends on the fluid velocity (Zhao et al., 2007; Szymczak and Ladd, 2009). The fluid velocity can be enhanced along permeable structures or zones in the rock, so that fractures, faults, grain boundaries and porous zones can favour flow and thus transport of chemical constituents. As advection is much faster than diffusion, it is much more effective in larger-scale systems such as large fossil fuel reservoirs or mineral deposits.

Reaction and advection/diffusion may influence each other. For example, they can be coupled in the sense that reactions may increase permeability causing a reactive infiltration instability (e.g. Chadam et al., 1986) where fluid-flow and hence further reaction is localized leading in the extreme case to "worm-holes" or caves in Karst systems (Szymczak and Ladd, 2009), replacement of relatively dense crystals through reaction-induced porosity development (Putnis and Putnis, 2007; Beaudoin et al., 2018) and infiltration of fluids and reactions into otherwise dry, impermeable systems (Jamtveit et al., 2000). Reactions may decrease permeability and arrest the reaction front propagation (Ruiz-Agudo et al., 2016). If reactions drive shrinkage and expansion, fracturing may occur, leading to additional pathways for advecting fluids (Ulven et al., 2014; Jamtveit et al., 2000). These positive feedback processes localize reactions and transport and drive faster material changes and strong localization. Reactive transport in reservoir rocks has been modelled extensively with an emphasis on the evolution of permeability (Saripalli et al., 2001; Zhao et al., 2007; Jamtveit et al., 2009; Chen et al., 2014; Kang et al., 2014; Mostaghimi et al., 2016). Methods range from smooth particle hydrodynamics to lattice Boltzmann methods and computational fluid dynamic techniques (Manwart et al., 2002; Tartakovsky and Meakin, 2006; Fredrich et al., 2006; Shabro et al., 2012; Chen et al., 2013). These studies show that there is a richness of complex interactions of fluid infiltration and reactions on permeability and porosity evolution in porous systems.

In a system where advection and diffusion are important, the dimensionless Péclet number (Pe) is used to describe the relationship between advection rate and diffusion rate for chemical transport as

$$Pe = \frac{vL}{D},$$  (eq. 1)

with $v$ the fluid velocity, $L$ the characteristic length scale of the system and $D$ the diffusion coefficient. At high *Pe,* advection is dominating, whereas at low *Pe* diffusion takes over. For example, the spacing and shape of wormholes in Karst systems changes as a function of the *Pe* number (Szymczak and Ladd, 2009). A fracture-dominated system where fluid infiltrates along the fractures would have a relatively high *Pe* number. In contrast, in a system where the fluid is stationary diffusion is dominant and its *Pe* number would therefore be low. *Pe* may change over time if the driving forces for the advection are changing, especially if the reaction changes the permeability (Ortoleva et al., 1987).

In a system, where reactions occur along with advection and diffusion, two additional dimensionless numbers are used in order to assess the influence of the relative rates of these processes. These two numbers are: (i) the Damköhler number I for reaction rate relative to advection rate

$$Da_I = \frac{R}{v},$$  (*eq. 2*)

and Damköhler number II to relate reaction rate relative to diffusion rate

$$Da_{II} = \frac{RL}{D},$$  (*eq. 3*)

with *R* the reaction rate. For example, Szymczak and Ladd (2009) show that both Damköhler numbers influence the shape and spacing of wormholes. A change in the Damköhler I number modifies the localization, width and spacing of wormholes. If the reaction is too fast, the localization of wormholes is hindered with the reaction front being smooth. In contrast, if the reaction is too slow the patterns become very fuzzy without developing wormholes (Szymczak and Ladd, 2009). Recent reactive transport simulations of reservoir rocks have also shown the importance of the Damköhler number for the alteration of pore space (Mostaghimi et al., 2016).

In this contribution we present a coupled numerical approach where reaction textures develop in a system that allows fluid advection, chemical diffusion and reaction to take place. We study the formation of reaction front patterns that develop in a simple granular aggregate with porous grain boundary regions representing a granular aggregate or breccia. We explore the phase-space between advection, diffusion and reaction rates to present a new classification of reaction front patterns and their link to the relative rates of the three main processes involved. Finally, we compare the numerical outcomes with natural examples and experiments of fluid-mediated replacement reactions, which emphasis that not only fluid pathways, but also the rate of reaction, have a major influence on reaction front patterns.

2. Numerical Set-up

2.1. General Model

We use a coupled hydro-dynamic model "Latte" within the microstructural modeling environment "ELLE" (Koehn et al., 2003; 2005; 2019; Bons et al., 2008; Sachau and Koehn, 2010; Sachau and Koehn, 2013; Ghani et al., 2013, 2015) and expand the model by adding advective and diffusive matter transport as well as a simplified iso-volumetric replacement reaction. We set-up the model to simulate the infiltration of a grain aggregate with more permeable grain boundaries and the progression of the reaction front where reactions are triggered by the presence of certain element concentrations that are changing due to advection, diffusion and reaction (fig. 2a). The numerical two-dimensional representation of a square slice of solid is represented by a triangular mesh of cells where clusters of cells make up grains. The run-cycle of the model starts with the initial granular geometry that defines the local porosity (fig. 2b). Fluid pressure and concentration are applied as boundary conditions. Note that the fluid pressure is ramping up linearly per time step. These are followed by a calculation of the infiltrating fluid represented by changes in fluid pressure and deriving the local Darcy velocity. The Darcy velocity is then used to calculate the advective matter flux followed by the diffusive flux. The new concentration of the reactant is finally used to drive the reaction and the local change in replacement is determined followed by a new cycle (fig. 2b). The granular aggregate has a porosity defined by the local solid fraction of the network with a background variation on the cell-scale and with grain boundaries having a higher porosity. The local permeability $K(\phi_{x,y})$ is calculated using the Carman-Kozeny relation (Carman, 1937; Ghani et al., 2013) according to

$$K(\phi_{x,y}) = \frac{r^2(\phi_{x,y})^3}{45(1-\phi_{x,y})^2} \qquad (eq\ 4)$$

where *r* is a fixed grain size and $\phi_{x,y}$ the local porosity. The fluid infiltrates the model realm from all four boundaries (fig. 2) where the fluid pressure is increased to initiate flux. These boundary conditions represent experiments of fluid infiltration into reactive samples in autoclaves with increased temperature, which we want to compare with the simulations. In natural settings high-pressure hydrothermal fluids will enter rocks from permeable fractures or faults. The fluid pressure evolution into the cell is derived using the following relation

$$\phi\beta\left[\frac{\partial P}{\partial t}\right] = \nabla \cdot \left[(1+\beta P)\frac{K}{\mu}\nabla P\right] \qquad (eq\ 5)$$

where $\phi$ is the porosity, $\beta$ the fluid compressibility, *P* the fluid pressure, *K* the permeability, $\mu$ the fluid viscosity. For a more detailed derivation see Ghani et al. (2013). For each time-step, equation 5 is used to calculate the fluid velocity $v$ from the Darcy flux $\phi\vec{v}$ for the advection of reactants according to

$$\vec{v} = -\frac{\frac{K}{\mu}\nabla P}{\phi} \qquad (eq\ 6)$$

In order to derive the transport of reactant into the system it is assumed that the four boundaries of the numerical model retain a constant concentration *C*. The different physical effects of advection, diffusion and reaction are separated (see eq. 9-11 below) and added after each time step according to

$$C^t = C^{t-1} + \delta C^t_{adv} + \delta C^t_{diff} + \delta C^t_{react} \qquad (eq.\ 7)$$

solving the general transport equation

$$\frac{\partial C}{\partial t} + \vec{v}\,\nabla C - D\Delta C = f. \qquad (eq.\ 8)$$

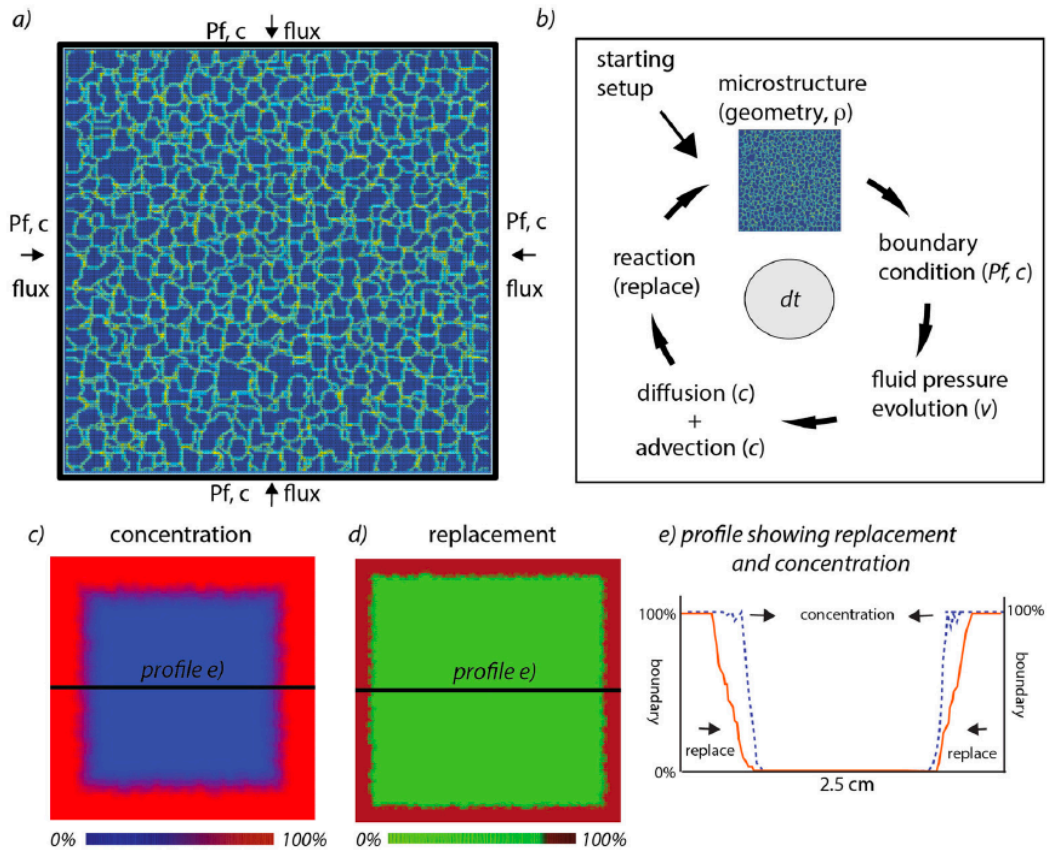

Figure 2. Illustration of the numerical model setup. a) 2D setup of the model where fluid and concentration are set at the boundaries with the concentration a constant and the fluid pressure increasing to produce a constant flux into the model. Grain boundary network with permeable grain boundaries appear in lighter colour. b) Numerical calculations loop in the model with the input from the grain aggregate (porosity/permeability) and the boundary pressure, then the fluid pressure evolution that gives the Darcy velocity followed by the mass transport equations related to advection and diffusion and finally the reaction. c) Concentration change into the model after a number of model runs; d) Related growth or mineral replacement patterns; e) 1D profile of the fraction of the maximum concentration (doted blue line) and fraction of complete replacement profile (solid orange line) at a given time, represented as a function of the particle position along a line passing through the middle of the simulations (reported on c, d).

The IMEX (IMplicite+Explicit; Asher et al., 1997) approach is used, where the advection is treated in an explicit and the diffusion in an implicit way with internal time loops in the advection to increase stability. This approach offers the possibility to study both, diffusion and advection dominated domains of the problem. Explicit in this case means solving the transport equation in a forward way in terms of time whereas the implicit solution of the diffusion equation uses a matrix inversion and solves the future time step at once. The advection is calculated in an explicit time-stepping method using the Lax-Wendroff scheme (Lax and Wendroff, 1960) according to

$$\delta C_{adv}^{t} = -dt\, \vec{v}\, \nabla C^{t-1} \qquad (eq.\ 9)$$

with $\phi\vec{v}$ the local Darcy velocity of the fluid determined from equation 6. The diffusion is calculated with an implicit method using the ADI algorithm (Alternating Direction Implicit, Bons et al., 2008) according to

$$\delta C_{diff}^{t} = dt\, \Delta\bigl(C^{t-1} + \delta C_{diff}^{t}\bigr). \qquad (eq.\ 10)$$

Finally, the reaction term in the equation calculates the reaction according to (Koehn et al., 2003)

$$R = k_r V_s \left(1 - \frac{C_a}{C_a^{eq}}\right), \qquad (eq.\ 11)$$

with the reaction rate $R$, $k_r$ a rate constant, $V_s$ the molecular volume of the solid, $C_a$ the concentration of $a$ and $C_a^{eq}$ the equilibrium concentration of $a$ in the fluid. Finally, the reaction rate is used to calculate the local replacement based on the existing replacement and the volume of particles. Particles that have been replaced by 100% become inactive.

For the sake of simplicity, the numerical model contains several underlying assumptions. The modelled reaction occurs at constant temperature, thus there is no exothermal or endothermal process active, and the reaction is iso-volumetric and does not affect the elastic properties nor the porosity of the material. Furthermore, the concentration in the fluid is thought to be sufficiently enough to lead to a replacement of the mineral. We assume that the permeability-porosity relation can be approached through equation 4 following a Carman-Kozeny relation, for the matrix as well as grain boundaries, which are thought to contain material and act as granular media as well. The diffusion constant is thought to be constant across the model irrespective of the porosity. Most of these assumptions can be changed in future models; however, this would complicate the interpretation, which is the reason why we currently use the most simplified setup for our study.

2.2. Set-up of simulations:

In all simulations the following parameters are used: dimensions of the solid 2.5x2.5 mm$^2$, porosity of the solid $\phi = 0.01 - 0.02$ (with the grain boundaries represented by zones of twice the porosity of the matrix), Carman-Kozeny grain size $r$ = 0.001 mm, fluid viscosity $\mu$ = 1.0x10$^{-3}$ Pa s, fluid compressibility $\beta$ = 4.5x10$^{-10}$ m$^2$/N, diffusion constant = 1.0x10$^{-10}$ m$^2$/s. For the reaction we vary the relative boundary concentration and use a reaction rate constant $k_r$ from 0.0001 to 0.01 mol/(m$^2$.s) and molecular volume $V_s$ = 0.00004 m$^3$/mol for calcite (Clark, 1966; Renard et al., 2004; Koehn et al., 2007), giving reaction rates of about 10$^{-6}$ to 10$^{-8}$ m/s. However, these values are only benchmark values and are rescaled to percentages in the plots such that 100% means full replacement for the reaction and 100% means full boundary concentration for the advecting and diffusing constituents. Models are run between 10000 to 100000 steps representing 2 minutes to about 10 hours, the time step for each model changes between 0.001 second to 0.3 seconds depending on the speed of the processes involved. The external pressure is ramped up simulating the heating of the fluid in the autoclave. This process takes a few minutes (between 2 and 3 minutes) and gives a second timescale to the pressure equation and influences the Darcy velocity and the advection.

The simulation is comprised of a complete infiltration of the material with fluid by increasing the fluid pressure at the boundaries leading to a continuous fluid flux into the model while the relative advection, diffusion and reaction rates are varied systematically allowing for a sensitivity analysis. The pressure was ramped up by 100 to 500 Pascal per time step up to pressures of 1 to 50 MPa representing the heated fluid in the autoclave.

2.3. Methods of analysis and representation

During the progression of the experiments the concentration in the fluid changes as a function of advection and diffusion and the composition of the mineralogy of the solid changes representing the exchange reaction. We show these changes in two ways, as 2-dimensional plots of the experiments showing the concentration change (fig. 2c) and the replacement reaction (fig. 2d, fig. 3) as well as profiles through the centre of the solid square recording both, concentration and replacement for single time steps (fig. 2e, fig. 4). Concentration changes in the 2-dimensional plots are shown in a linear colour-scale between blue (0%) and red (100%) and the replacement reaction is shown in a stepped colour scale between green (0-85%) and brown (85– 00%) to visualize the reaction front morphology. The reaction front morphology is described as smooth, irregular, rough and replaced as a function of the amplitude of the boundary roughness relative to the average grain size representing the wavelength of the signal (fig. 3). If the amplitude/wavelength ratio is below 0.5 the roughness is defined as smooth, if the ratio is between 0.5-1.5 it is defined as irregular, and if it is above 1.5 the boundary is defined as rough. If the reaction front is absent, i.e. it runs across the whole aggregate, the pattern is referred to as replaced. We characterize the transport using the Péclet number (eq. 1), with high Péclet representing advection-dominated fluid infiltration, and low Péclet representing a more diffusion-dominated system. The reaction is first represented by the reaction rate (so that it is independent of transport) and plotted versus the Péclet number (figs. 5, 6). In a final plot of non-dimensional phase-space the Damköhler I number (eq. 2) representing the relative advection to reaction rate is plotted versus the Péclet number (fig. 7).

3. Results

3.1 Concentration and replacement pattern-development through time

Figure 3 shows the replacement of grains in the mineral aggregate through time for three example simulations developing a smooth, irregular and rough reaction front. The brown dark-colour represents a high percentage of new mineral growth whereas green represents a low percentage of new mineral growth. The first simulations (fig. 3a) show a reaction with a slow advection represented by a Péclet number of 10 where diffusion becomes important. The reaction rate is fast and the corresponding Damköhler I number $10^{-3}$. The second experiment (fig. 3b) shows a reaction with medium to fast advection rate with a Péclet number of 75 and a fast growth with a Damköhler I number of $10^{-3}$. The third experiment (fig. 3c) shows a reaction with a high Péclet number of 100 and a slow growth with a Damköhler I number of $10^{-4}$. The three experiments (fig. 3a-c) show different timescales of reaction front progression and distinctly different reaction front roughness. The front in the first experiment (fig. 3a, slow advection, fast growth) is smooth, the front progresses relatively slowly over 20 minutes into the simulation box, while the reaction front becomes smeared out, i.e. the width of the mixed reacted and unreacted material (green area in Fig. 3a) increases, and the corners of the reaction front are rounded. The grain boundaries cannot be seen signifying that there is no preferred reaction along grain boundaries. The front in the second experiment (fig. 3b, medium-fast advection, fast growth) is irregular on the scale of single grains, as the reaction front *enters* the grain boundaries, i.e. there is a clear preference of reaction along grain boundaries. The reaction front becomes visible at the boundaries after about 1.7 minutes and is then filling out most of the box within 2-3 minutes. The front in the third experiment (fig. 3c, fast advection, slow growth) is rough on the scale of several grains where the grain boundaries are marked by reaction products. The reaction front enters the simulation box after about 50 minutes and then fills out most of the box where the grain boundary infiltration front is followed by an outer rim of fully reacted material. All three simulations show a reaction front morphology that is not changing over time.

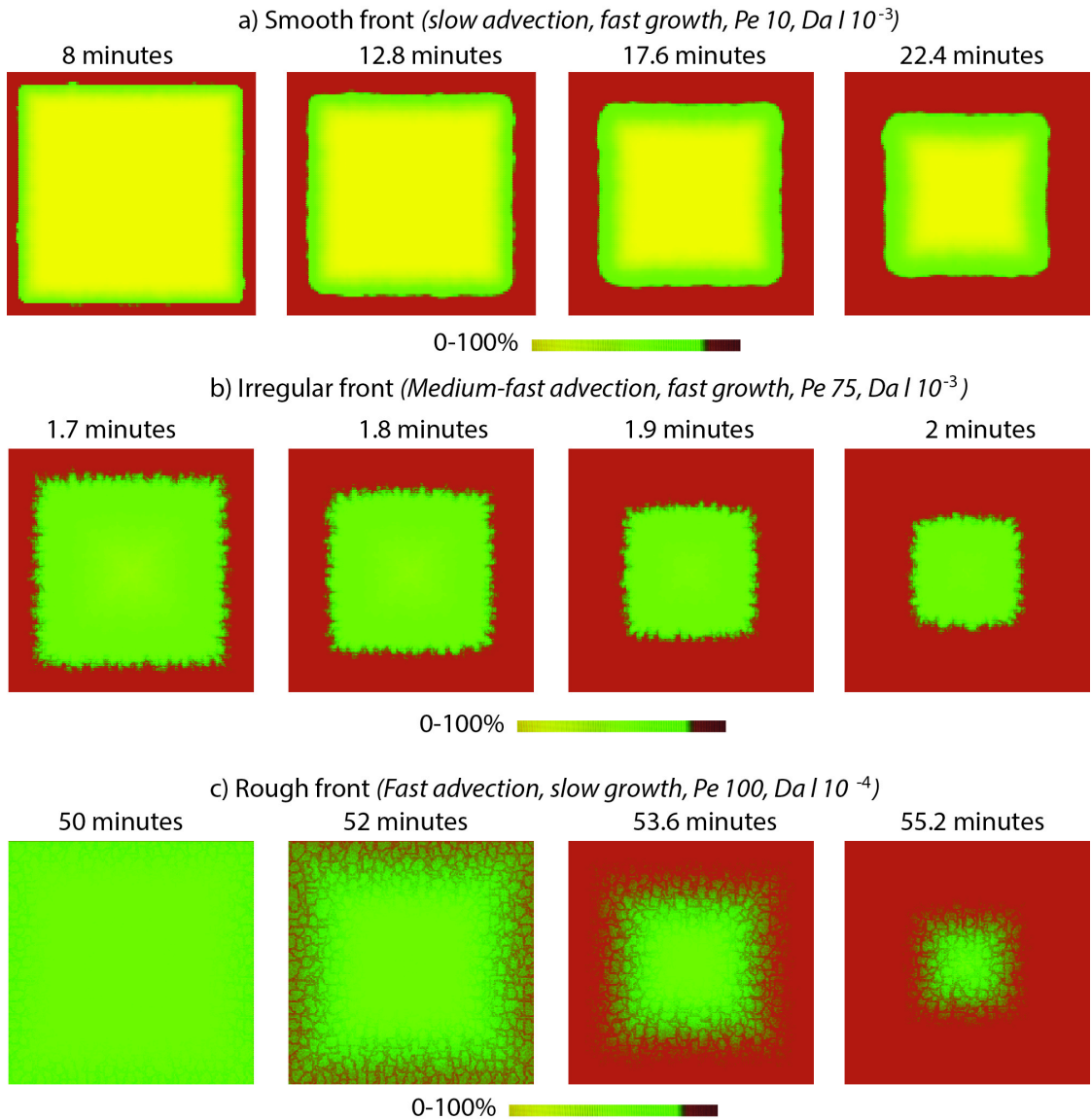

Figure 3. Three simulations with variably rough reaction fronts and images showing the mineral growth over time. a) Simulation 1 shows a slow advection where diffusion becomes important producing smooth interfaces and rounded corners. b) Simulation 2 shows a medium advection and a fast growth so that the developing structures are irregular on the grain scale. c) Simulation 3 shows a faster advection and much slower growth so that the final growth features accentuate the grain boundaries in a relatively wide zone and the reaction front is rough.

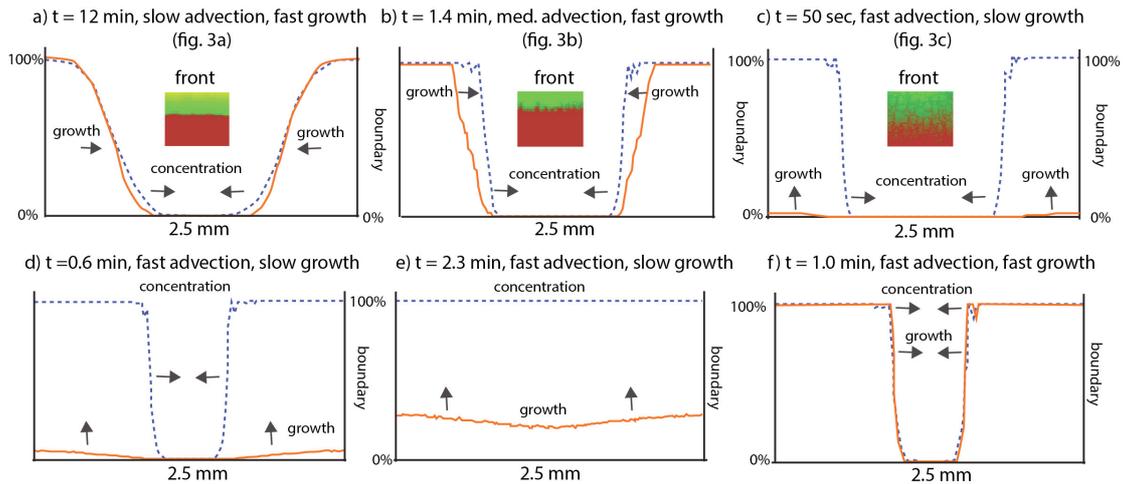

Figure 4. 1D profiles of the relative infiltration of the fluid concentration (fraction of maximum concentration) into the model (dotted blue line) and the following growth front (solid orange line, fraction of complete volume reacted) following a horizontal line crossing the model as illustrated on fig. 2c, d. a-f profiles relate to a number of different models at variable time steps. a-c) Fast growth that keeps up either with advection or diffusion and thus mainly covers the incoming front. d-f) Difference between the advective front coming in fast followed by a very slow reaction front with a small slope inwards that can capture and enhance the grain boundary network.

Figure 4 illustrates the difference between a concentration and a reaction profile through the different experiments, with the profiles running along the x-axis and through the centre of the simulation box (fig. 2). The dashed blue line shows the relative fluid concentration infiltrating the sample, whereas the solid orange line shows the relative growth of the new mineral (or the replacement), and both are scaled to 100%. Figure 4a show an experiment with slow advection and fast growth similar to experiment I in figure 3a, figure 4b shows an experiment with medium advection and fast growth similar to experiment II in figure 3b and figure 4c shows an experiment with fast advection but slow growth similar to experiment II in figure 3c. Figure 4a shows an example that is diffusion dominated and has a fast growth (relative to diffusion timescales). Diffusion leads to a smooth reaction front that is blurred but still relatively narrow (see reaction front in inset). The simulation with fast advection and fast growth on the other hand (fig 4b) shows a steep infiltration gradient of the concentration that is followed by a similarly steep gradient of the growth front. Therefore, these experiments show a thin, steep front (see reaction front in inset) that infiltrates the sample and is irregular on the grain-scale. Simulations with fast advection and slow growth (fig. 4c) show an advection dominated step, where the fluid is infiltrating the material with some roughness at the infiltration front due to fingering and grain boundary infiltration. This infiltration of concentration is followed by a relatively slow growth with a minor gradient into the sample. This slow growth results into an apparent preferential growth along the grain boundaries over time and shows a very shallow slope as a relic of the initial infiltration front (figs. 4d,e). Therefore, the resulting pattern will show a relatively wide zone where the grain boundaries become visible and the front is very rough and infiltration and growth are almost completely coupled (see inset in fig. 4c). Finally, a very fast reaction traces the advection completely and therefore has a very steep front and leads to a complete reaction without minor porosity and thus infiltration variations (fig. 4f).

3.2. Scaling of reaction front pattern as a function of Peclet number, reaction rates and Damköhler number

In order to illustrate the different reaction front patterns that develop in the different advection-diffusion-reaction scenarios we first plot a matrix of experiments in a diagram of Péclet number versus the reaction rate on a broad (fig. 5) and then on a more detailed scale (fig. 6) and finally show the patterns of the two dimensionless numbers Péclet versus Damköhler I in phase-space (fig. 7). The extreme variation of patterns on a broad-scale is shown in figure 5 where a low Péclet number of 1 produces diffusion dominated rounded, smooth and relatively sharp reaction fronts.

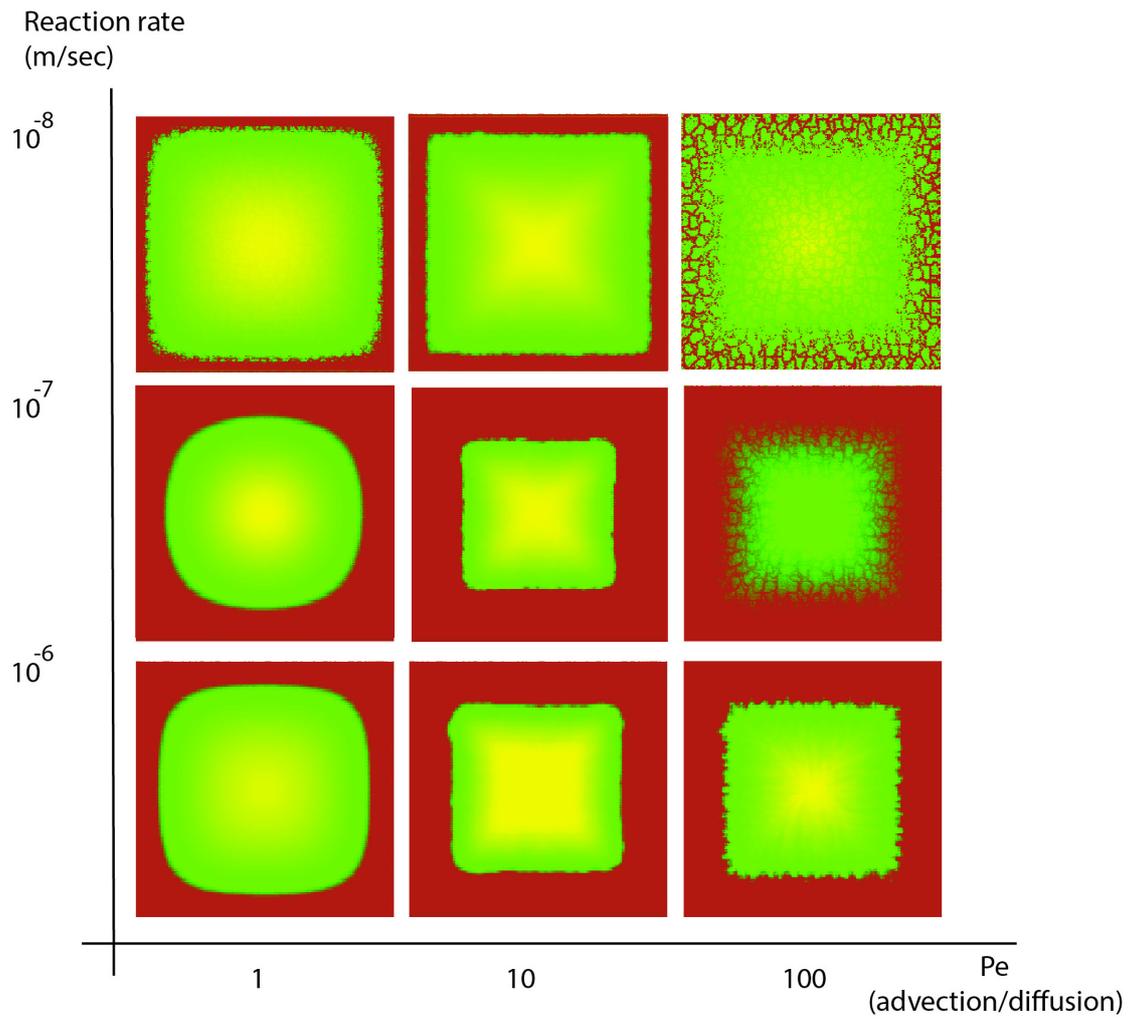

Figure 5. The developing reaction patterns on a rough scale in Péclet number versus reaction rate space. Rough reaction interfaces develop towards high Péclet numbers and slow reactions. Smooth and progressively rounded patterns develop in the low Péclet number domain where diffusion is dominating.

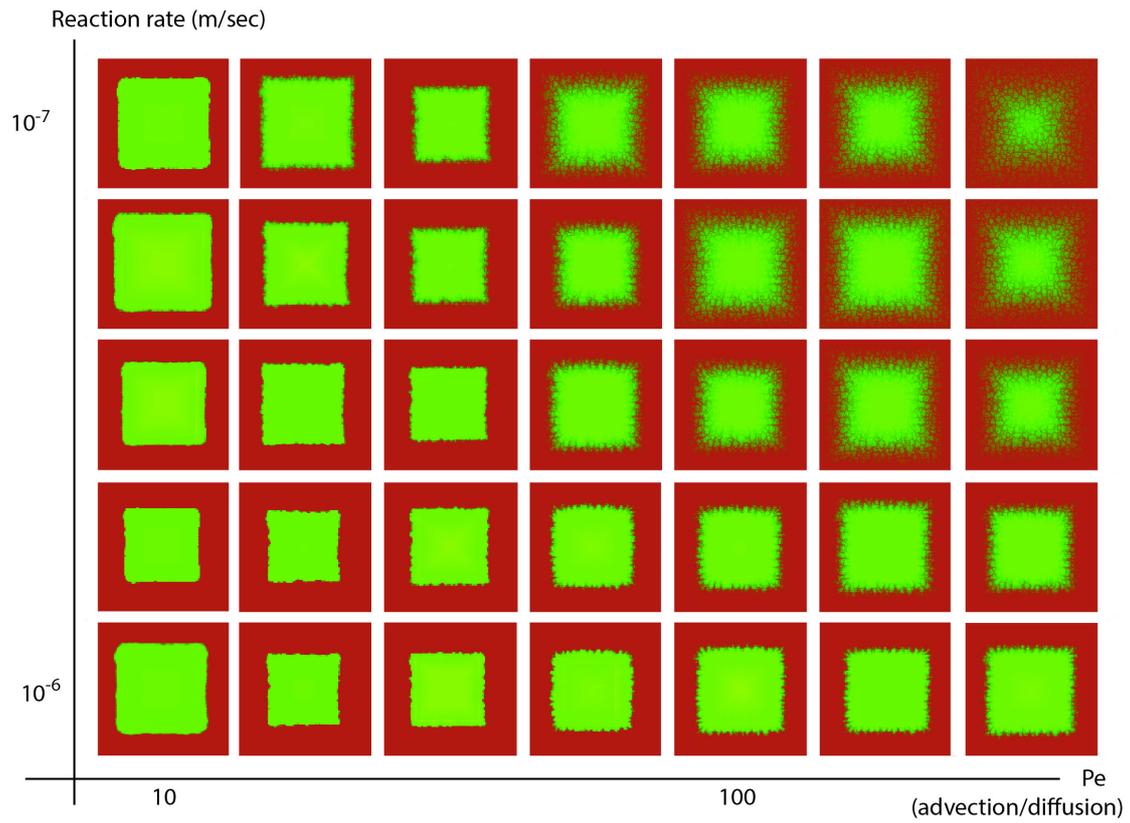

Figure 6. Magnified and more detailed version of fig. 5 at the transition between smooth and rough reaction interfaces. The matrix shows that an increase in Péclet number in general leads to an increase in the roughness with an important transition after a Péclet number of about 50. However the figure also illustrates that in order to develop a reaction front with significant roughness that is on a larger scale than the grains the reaction needs to be slower than about $5 \times 10^{-7}$ m/s.

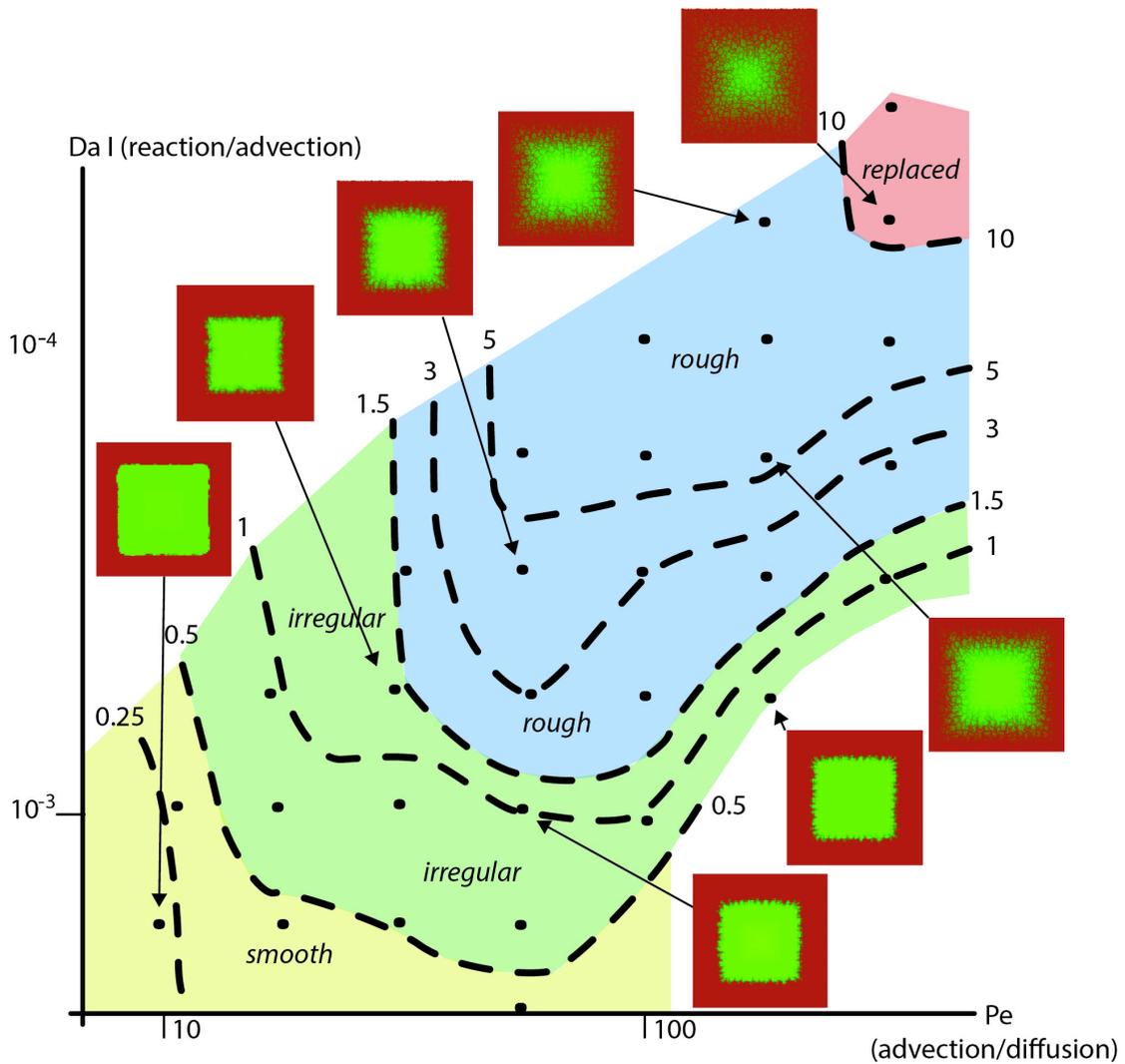

Figure 7. Evolution of the different reaction patterns in non-dimensional phase space of Péclet versus Damköhler I numbers. The roughness is defined by the amplitude/wavelength ratio of the signal (number next to dashed lines). Four main areas of patterns can be seen, smooth interfaces at the lower left hand corner with low Péclet and high Damköhler I numbers (yellow zone). This zone is followed by a zone with irregular interfaces that curves from low Damköhler I numbers down and then up again towards high Péclet numbers (green zone). Rough interfaces on the scale of several grains are shown in the blue zone at high Péclet and low Damköhler I and a full replacement with rough growth across more than 10 grains is shown in red at the uppermost corner of the diagram. Data points are shown as black dots with example images of interfaces.

For a very slow reaction rate of $10^{-8}$ m/s the pattern becomes rough on a very small scale driven by the reaction. Towards Péclet numbers around 10, reaction front propagation becomes more advection dominated resulting in a smooth, at corners rounded reaction front. At Péclet numbers around 100 advection dominates and the reaction front becomes irregular to rough. At fast reaction rates of $10^{-6}$ m/s the front is irregular on the grain scale but towards slower reaction rates of $10^{-7}$ and $10^{-8}$ m/s the front is rough, where grain boundaries start to show up and dominate the pattern. Figure 7 shows a more detailed matrix of figure 6 illustrating the transition from smooth through irregular and rough reaction front patterns. The matrix clearly illustrates that increasing Péclet numbers from 10 towards 100 increases the irregularity of the front. However, especially at higher Péclet numbers the reaction rate becomes also important with fast reaction rates ($10^{-6}$ m/s) producing a front with irregularities on the grain-scale, whereas slow reaction rates ($10^{-7}$ m/s) lead to rough fronts with grain boundaries showing up in the reaction. The most extreme infiltration takes place when the Péclet number is high and the reaction rate is low. The variation of the pattern and the changing roughness of the interface can be illustrated in phase-space of the two dimensionless numbers Péclet and Damköhler I (fig. 7). Note that the Damköhler I number on the vertical axis is plotted from high to low numbers to compare with literature data (Szymczak and Ladd, 2009). In the lower left hand corner of the diagram at low Péclet and high Damköhler I numbers, the reaction front is smooth (amplitude/wavelength ratio < 0.5). At lower Damköhler I accompanied by higher Péclet numbers, the reaction front becomes irregular on the single grain-scale (ratio between 0.5-1.5). The boundary between a smooth and irregular front is almost diagonal across the diagram. The zone where the reaction front is irregular curves around from low Damköhler I to high Péclet numbers. Towards the upper right-hand corner of the diagram the pattern becomes rough (ratio > 1.5) and is dominated by multi-grain boundary infiltration. In the uppermost right-hand corner of the diagram (ratio > 10) complete infiltration or replacement occurs. The phase-boundaries illustrate an increase of roughness with increasing Péclet number from Péclet 10 to 100 but then a decrease in roughness towards higher Péclet numbers controlled by the Damköhler I number.

4. Discussion

## 4.1. General model behaviour

Our numerical simulations illustrate different scenarios that produce rough, irregular and smooth reaction fronts. In the most extreme cases of "roughness" all grain boundaries in the aggregate are marked by reaction products, a pattern that is very similar to replacement reactions in fossils, sedimentary basins and metamorphic terrains. Advection is the main driving force for fluid infiltration into the system and for the development of roughness due to more permeable grain boundaries and advection fingering (Jonas et al., 2014; Kar et al., 2015; Plümper et al., 2017; Beaudoin et al., 2018). Advection, however, is not always enough to produce very rough fronts. In the case of complete infiltration the roughness only develops significantly if the reaction is slow. This is related to the fact that a fast reaction will follow the infiltration front and will only be able to superimpose the local anisotropic advection on the grain-scale or on the scale of advective fingers. However, if the reaction is slow, the mineral growth front into the material does not represent the advection front moving inwards, but rather the anisotropy of the grain boundary infiltration. The growth has a memory effect of the advective infiltration and preserves this pattern when it slowly replaces the whole aggregate. We envisage that this scenario can produce replacement of large bodies, by infiltrating them through advection followed by a slow growth that preserves heterogeneity of the rocks, even though the material is replaced. In this case the reaction does show the differential permeability of the rock but not the actual fluid infiltration, even though advection is still needed in order to attain a memory effect of the rock fabric. A slow reaction after infiltration also means that if the reaction fills pore-space and reduces the permeability, it is not clogging the fluid pathway, at least for one single infiltration event as modelled here. However, there is no reason why multiple fluid infiltration events in a cyclic manner into the material do not produce very similar structures to our simulations. The simulations have several timescales defined by the pressure diffusion equation, the external boundary condition of ramping up the pressure, the associated fluid velocity and the diffusion timescale. Diffusion-time scales as a function of the diffusion coefficient divided by the length scale squared. For the pressure equation this gives a time of roughly one second to diffuse the pressure into the experimental sample of 2.5 mm. In this case the external boundary condition of ramping up the pressure becomes important, because every increase in pressure leads to a new fluid pressure diffusion into the sample. The pressure ramping up takes 2 to 3 minutes and this timescale is then controlling the flux of material into the sample. In this case the fluid velocity from equation 6 gives a velocity of $10^{-6}$ mm/sec for the initial advection. Since the fluid velocity is dependent on the pressure gradient this velocity goes up linearly with an increase of the pressure at the boundary, if this increase is faster than the pressure diffusion timescale, which it is in the simulations. A pressure gradient of 1MPa/model-unit then leads to a fast infiltration of matter through advection filling the box within one minute. The pressure diffusion timescale becomes much smaller for larger systems with a time scale of 18 seconds for a cm size domain, 50 hours for a meter size domain and 60 years for a reservoir-type domain of 100m. The matter-diffusion timescale into the system is in the order of 7min for a domain of 0.2mm and 44 days for the whole experimental domain of 2.5mm.

4.2. Comparison of results to experimental replacement reactions

In this section we compare the numerical simulations with results from replacement experiments. In many replacement-reaction experiments the setup is similar to the numerical setup presented here, where a square piece of material is exposed by a reactive fluid at all four sides and the replacement reaction is monitored through time. Here we present two sets of replacement experiments in which polycrystalline Carrara marble (pure white marble from Carrara, Italy, 99.7% calcite, average grain size diameter of 100 μm) cut into regular cubes (2-3 mm), is replaced by calcium phosphates. Experiments were performed following previously published protocols (Kasioptas et al., 2008; Pedrosa et al., 2016) where samples are immersed into a reactive fluid and inserted into a hydrothermal autoclave at temperatures of 180ºC.

Only the reaction rate of the replacement varied by reacting the marble with either (set 1) fluorine-containing phosphate solutions (1.0 M $(NH_4)_2HPO_4$ + 0.1 M $NH_4F$) or (set 2) sodium chloride-containing phosphate solutions (1.0 M $(NH_4)_2HPO_4$ + 0.5 M NaCl). Set 1 shows a reaction that proceeds normal to the outer perimeter of the sample with fast reaction rates, the sample being half replaced in about 5 days and the interface being rough on the grain scale (fig. 8a). Set 2 experiments show a slow-moving reaction with only 10% of the sample replaced after 15 days and the reaction being primarily associated with grain boundaries and fractures in the sample (fig. 8c). Figure 8a and c show the experiments of Set 1 (fig. 8a) and Set 2 after 5 and 15 days respectively (fig. 8c) compared to two simulations (fig. 8b and d).

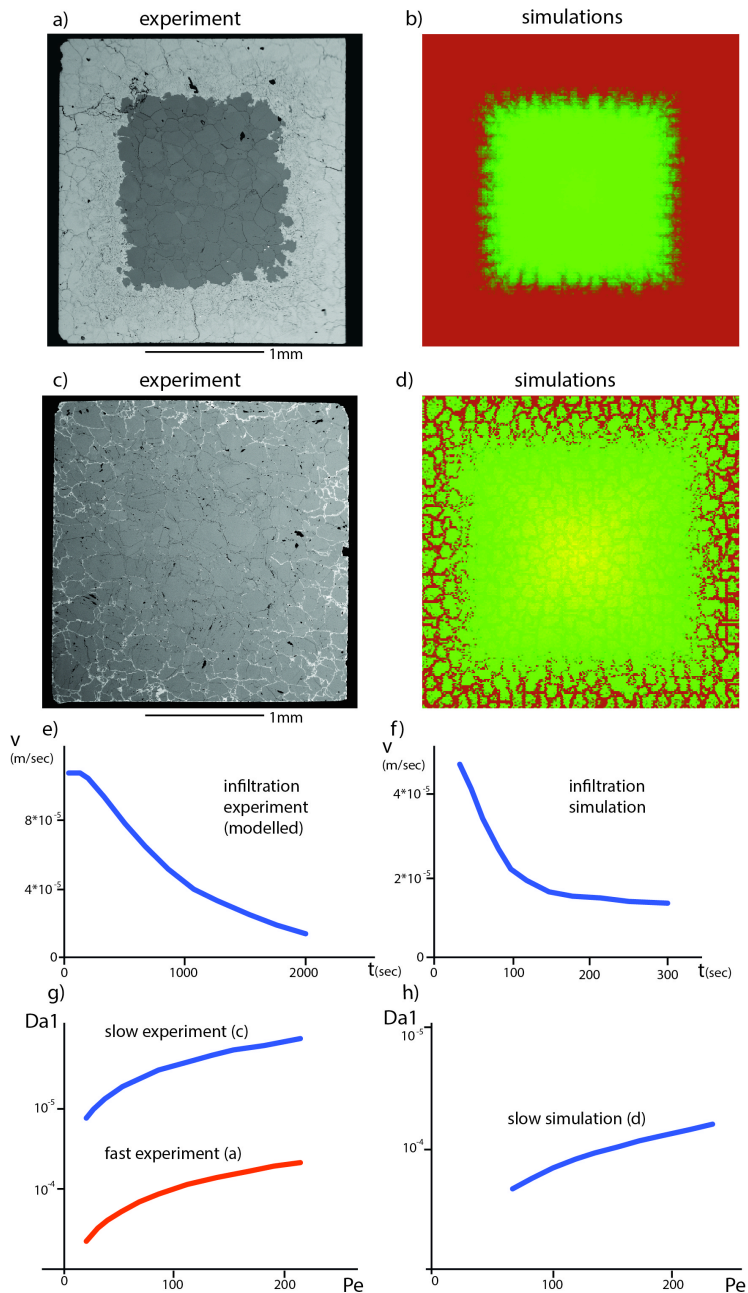

Figure 8. Experimentally produced patterns versus the simulations. a) Back-scattered SEM (Scanning Electron Microscope) image of a cross section of a Carrara marble sample after a replacement experiment where the fluid infiltrates the material from the sides and the reaction takes place parallel to the boundaries with a minor roughness that develops on the grain boundary scale. Unreacted calcite grains are grey, reaction products are white. b) Simulation that produces the same pattern at a Péclet number between 75 and 500 and a Damköhler I number of $10^{-3}$. In this case the high Péclet number indicates that the transport is advection dominated whereas the Damköhler number indicates that the reaction is relatively fast. c) Back scatter SEM image of a cross section of a Carrara marble sample after a replacement experiment where the fluid infiltrates from the lower and the left hand side boundaries (the figure shows the lower left hand corner of an experiment with a similar setup to a)). The reaction takes place mainly along the grain boundaries. Unreacted calcite grains are grey, reaction products are white. d) Simulation with a similar setting to c where the grain boundaries react. In this case the Péclet number is about 500 and the Damköhler I number $10^{-4}$.

The transport is advection dominated but the reaction is relatively slow. e) Modelled infiltration velocity for the experiments as a function of time (see text for derivation). f) Infiltration in a simulation determined from the infiltration front progression over time. g) Plot showing Damköhler versus Péclet number for the two experiments. Both experiments show a decrease in Péclet and an increase in Damköhler numbers over time. h) Damköhler versus Péclet number for a slow simulation showing the same decrease in Péclet over time that can be observed in the experiments.

The simulations mimic the patterns of the replacement reactions very well with simulation shown in figure 8b showing a small roughness on the grain boundary scale whereas the simulation shown in figure 8d shows infiltration along grain boundaries along the rim of the experimental charge. The settings for the simulation shown in figure 8b is set to model fast reaction rate, and fast advection and fast growth (Fig. 6) resulting in a rough front on the grain scale whereas the simulation shown in figure 8d is set to model slow reaction rate coupled with fast advection relative to slow growth i.e. high Pe number resulting in strong grain boundary infiltration (fig. 5). One has to note however, that the time scales in the numerical simulations and the experiments are not the same with the experiments taking longer (days) than the simulations (minutes to hours). This discrepancy may be present due to the lack of data on the exact setting of the experiments in terms of external pressure, temperature gradients, reaction rates as well as diffusion constants. The experiments are still within a time frame where diffusion is only present at small scales, and the slow reaction of experiment two could be reproduced in the model. However, experiment one with sharp reaction front has a faster advective timescale in the simulations in the order of a couple of minutes driven by ramping fluid pressure boundary. An additional parameter may slow down this advection in the experiments, potentially through a reaction that changes the porosity. An alternative advective transport mechanism that is not pressure-driven is chemically-driven convective-flow into dead-end pores through transient diffusioosmosis (Kar et al., 2015). This process leads to fluid velocities of 10 to 50 micrometers/second, so that fluid could infiltrate the experimental setup within minutes. Even though this driving mechanism is different, the resulting patterns should look similar to those of our simulations. However, the timescales may vary.

In order to compare the experiments better with the simulations we estimate the infiltration velocity into the experimental samples, calculate Peclet and Damköhler numbers and compare them to simulations (fig. 8e-h). We model the infiltration velocity by considering temperature diffusion into the sample as a function of the temperature of the autoclave (180ºC) and the temperature in the sample (20ºC) using a simple one-dimensional finite difference approach. The temperature is then used to calculate the local fluid pressure that progresses into the sample and the pressure gradient from the boundary towards the centre using the bulk Modulus of water $G$ (2 GPa) and the coefficient of thermal expansion $\alpha$ (0.00006 K$^{-1}$) as $\delta P = \delta T \frac{\alpha}{1/k}$ . The pressure gradient is then used to determine the fluid velocity or infiltration velocity using equation 6 (fig. 8e). We then can derive the Peclet number for the experiments and using the time scale of the reaction from Figure 8b,d we can determine the Damköhler number as well (fig. 8g). For the simulations we use the progression of the infiltration front as a function of time to determine the velocity directly (fig. 8f). The velocity is then also used to determine the Peclet number and the velocity of the reaction front can be used to determine the Dahmköhler number (fig. 8h). This gives us a direct comparison of the experiments and the simulations with the fluid infiltration decaying in both examples where the experimental infiltration velocity seems to be higher. In addition the infiltration slows down completely in the simulations after 200 seconds whereas the infiltration takes longer in the experiments. However, the overall behaviour is similar with both simulations and experiments having an infiltration velocity that decays over time as a function of the fast driving processes in the beginning followed by a diffusion like decay. The plots of Damköhler versus Peclet number show a very similar trend between experiment and simulation with both showing a decay in Peclet number as a function of the slowing down infiltration velocity and an increase in Damköhler number because of this. The Damköhler numbers of the slow simulation are not as low as those of the slow experiment illustrating the mentioned fact that the growth in the experiments is slower than the growth in the simulations.

4.3. Applicability of results to natural examples and use of reaction front pattern in determining relative rates of reaction, advection and diffusion

Our experiments show that reaction patterns in rocks can be used as a toolbox to understand paleo-reaction and transport rates. For example, figure 1a shows a natural example of a reaction around a fracture in a sandstone where $Fe^{3+}$ is reduced to $Fe^{2+}$. The pattern is frozen in time and shows an irregularly rough front, either with a low Peclet number of around 10 or a higher Peclet number and a higher Damköhler number I of around $10^{-3}$. Figure 1b shows a dolomitization front with the darker lower part of the rock being dolomitized with low porosity whereas the upper part has a high porosity and is not dolomitized. The front is rough and mimics a fluid infiltration that is frozen into the rock record by the reaction. The infiltration into the rock seems to be more driven by fractures/sedimentary layer boundaries than grain boundaries. According to our study this pattern needs a high Peclet number so that advection dominates the transport mechanism and an intermediate to low Damköhler number so that the anisotropy in permeability along for example grain boundaries or fractures is preserved by the reaction. Figure 1c shows a naturally occurring dolomitization front (brownish in outcrop) within a carbonate. The reaction front is rough on the scale of several grains. This pattern would need a Peclet number larger than 100 and a low Damköhler number in the range of $10^{-4}$ (Fig. 7).

The Damköhler number also influences the alteration of pore-space in reservoir rocks where the rocks dissolve homogeneously at a high Peclet and low Damköhler I number similar to patterns that we see in our reactions (Mostaghimi et al., 2016). A difference occurs at high Damköhler numbers that leads to the development of large channels and a high porosity in the model of Mostaghimi et al. (2016). This is also in contrast to wormhole formation, where a high Damköhler I number hinders the localization and thus formation of wormholes with the front being smooth (Szymczak and Ladd, 2009), which is similar to our patterns. In summary it is important to notice that what we see in the rock record as reaction patterns is a function of both, the rates and mode of transport and reaction. The presented work shows the complexity of the interplay of these processes. Coupling the influence of both processes to a dynamic porosity promises to provide a toolbox that can be used as a deductive and predictive tool for (paleo-) fluid-flow and reaction, reservoir evolution, ore body formation and in general fluid-rock interaction in the Earth's crust.

5. Conclusion

In this contribution we modelled the infiltration of fluid into a small rock sample (2.5 x 2.5 mm) with permeable grain boundaries, with mass transport as a function of advection and diffusion and a consecutive reaction. Advection-dominated infiltration produces irregular to rough boundaries with grain boundary infiltration and fingering, whereas diffusion-dominated transport favours smooth boundaries. In addition, the rate of the reaction relative to the fluid infiltration process is crucial where fast reaction produces reaction fronts that are smooth or only irregular on the grain boundary scale whereas slow reactions memorize the anisotropy of the infiltration process, develop rough interfaces and produce a grain boundary network. These patterns can also be illustrated in non-dimensional phase-space using the Péclet and Damköhler I numbers, with smooth fronts at low Péclet and irregular to rough fronts at high Péclet and low Damköhler numbers. In the extreme case at very high Péclet and low Damköhler numbers the complete grain boundary network can be reproduced by the reaction in a manner that is reflecting a replacement process. Our study indicates that a dominating advection process as well as a slow reaction are important for rough fronts. We show that our results mimic patterns found in experiments and in nature and argue that replacement reactions of large areas that preserve the initial rock texture (fossils, sedimentary or crystalline structures) may be driven by initial advection of fluids into the system followed by a slow reaction that "freezes" the initial pattern. Our results indicate that what we observe as patterns in rocks is not only a function of transport mechanisms but also and importantly its dynamic interplay with reaction and reaction rates.

Data availability

The simulation input and output data used to support the findings of this study are available from the corresponding author upon request. The basic software for the simulations can be found and downloaded at http://elle.ws and the corresponding author will make the additional code available upon request.

Acknowledgements

The authors acknowledge suggestions by two reviewers on an earlier version of the manuscript that helped to improve the work. The authors also acknowledge support of the ITN FlowTrans, a funding from the European Union's Seventh Framework Programme for research under grant agreement no 316889. NEB is funded through Isite E2S-ANR PIA-RNA. RT acknowledges the support of INSU, ALEAS program, of the IRP France-Norway D-FFRACT, and of the Research Council of Norway through its Centres of Excellence funding scheme, project number 262644.